\documentclass[aps,pra,amsthm,amsmath,twocolumn,superscriptaddress,showpacs,nofootinbib]{revtex4}

\usepackage{amsthm}
\usepackage{color}

\definecolor{gray}{gray}{.5}

\input{Qcircuit}

\renewcommand{\Qcircuit}[1][0em]{\xymatrix @*[o] @*=<#1>}
\renewcommand{\node}[2][]{{\begin{array}{c} \raisebox{.6em}{$\scriptstyle#1$} \\ {#2} \\ \rule[-.3em]{0em}{0em} \end{array}}\drop\frm{o} }
\newcommand{\nodeDown}[2][]{{\begin{array}{c} \rule[.9em]{0em}{0em} \\ {#2} \\ \raisebox{-.35em}{$\scriptstyle#1$}\end{array}}\drop\frm{o} }

\newcommand{\nghb}{{\mathcal N}}

\usepackage[sort&compress]{natbib}

\begin{document}

\title{Modeling Pauli measurements on graph states\\
with nearest-neighbor classical communication}

\author{Jonathan Barrett}
\email{jbarrett@perimeterinstitute.ca}\affiliation{Perimeter
Institute for Theoretical Physics,
31~Caroline Street N, Waterloo, Ontario N2L 2Y5, Canada}

\author{Carlton M.~Caves}
\affiliation{Department of Physics and Astronomy, University of New Mexico,
Albuquerque, NM 87131}

\author{Bryan Eastin}
\affiliation{Department of Physics and Astronomy, University of New Mexico,
Albuquerque, NM 87131}

\author{Matthew B.~Elliott}
\email{mabellio@unm.edu}
\affiliation{Department of Physics and Astronomy, University of New Mexico,
Albuquerque, NM 87131}

\author{Stefano Pironio}
\email{Stefano.Pironio@icfo.es} \affiliation{ICFO-Institut de Ci\`encies Fot\`oniques, Mediterranean Technology Park, 08860 Castelldefels (Barcelona), Spain\\
Institute for Quantum Information, California Institute of
Technology, Pasadena, CA 91125}

\begin{abstract}
We propose a communication-assisted local-hidden-variable model that
yields the correct outcome for the measurement of any product of Pauli
operators on an arbitrary graph state, i.e., that yields the correct
global correlation among the individual measurements in the Pauli product.
Within this model, communication is restricted to a single round of
message passing between adjacent nodes of the graph.  We show that any
model sharing some general properties with our own is incapable, for at
least some graph states, of reproducing the expected correlations among
all subsets of the individual measurements. The ability to reproduce all such correlations is found to depend on both the communication distance and the symmetries of the communication protocol.
\end{abstract}


\maketitle

\section{Introduction}

Graph states are multi-partite entangled states that play many important
roles in quantum information theory.  The class of graph states is
equivalent, by local unitaries in the Clifford group, to the class of
states stabilized by Pauli operators \cite{schlingemann,vandennest}.  This
class includes Bell states, GHZ states, basis states for stabilizer codes,
cluster states, and many others.  Of particular interest are the cluster
states, which are the graph states represented by two-dimensional square
lattices \cite{briegel}.  Cluster states have been shown to be sufficient to allow
universal quantum computation within a measurement-based
scheme~\cite{oneway}.  For this reason, a complete understanding of the
entanglement properties of graph states would likely improve our
understanding of the role entanglement plays in quantum computation, as
well as teaching us about some of the most useful states in quantum
information theory. Graph states and their applications are reviewed in Ref.~\cite{hein}.

Both G\"{u}hne {\it et al.}~\cite{guhne} and Scarani {\it et
al.}~\cite{scarani} have shown that graph states display nonlocal
properties under the measurement of Pauli operators.  In this work, we
further our understanding of the nonlocality of graph states by
introducing a communication-assisted local-hidden-variable (LHV) model
that predicts the outcome of measuring an arbitrary Pauli product on an
arbitrary graph state.  Since graph states violate Bell-type inequalities,
the model necessarily involves communication.

Our investigation is inspired by that of Tessier {\it et
al.}~\cite{tessier:lhv}, who described a communication-assisted LHV
model for arbitrary Pauli measurements on a GHZ state.  In the spirit
of Tessier {\it et al.}, we formulate our LHV model in terms of hidden
variables that can be thought of as specifying values for the $x$, $y$,
and $z$ spin components of the qubits.  In general, a communication
protocol might permit the party at a particular qubit to communicate to
any other party what Pauli measurement is made on its qubit.  In our
communication protocol, however, we restrict communication to be
between parties corresponding to nodes that are adjacent in the
underlying graph.  This restriction to communication only with
neighbors in the graph makes intuitive sense if we think of a graph as
a recipe for constructing the corresponding graph state.  In that case,
nodes that are connected have interacted in the past and therefore
occupy a privileged position with regard to exchange of information.
We call a protocol that restricts communication to neighbors a {\it
nearest-neighbor communication protocol}.

Although our communication-assisted LHV model predicts correctly the
outcome of the measurement of any Pauli product, it fails in some cases to
predict the expected correlations for subsets of the individual
measurements in a Pauli product.  By considering restricted classes of
graphs, we show that two general properties of our model assure its
failure.  Perhaps unsurprisingly, one of these is the limitation to
nearest-neighbor communication.  More generally, we consider protocols
with a limited \textit{communication distance}, defined as the number of
successive edges through which information can be sent, and we show that
any protocol whose communication distance is constant or scales less than
linearly with the number of qubits fails to predict some submeasurements
correctly.  Less obvious is a second problem of our protocol, which we
call \textit{site invariance}, i.e., the property that nodes in symmetric
situations perform the same action.  We consider the effects of each of
these properties in some detail and show that if a protocol has either
property, it fails on some submeasurements.

This paper is organized as follows. In Sec.~\ref{sec:states} we introduce
the formal definition of graph states.  In Sec.~\ref{sec:model} we
describe our model and prove that it correctly predicts the global result
of any Pauli measurement on a graph state, i.e., predicts the global
correlation among the individual measurements in the Pauli product.   In
Sec.~\ref{sec:failure} we demonstrate that neither site-invariance nor any
fixed communication distance is compatible with the goal of reproducing
all subcorrelations, though we do demonstrate that a site-invariant
protocol can reproduce all subcorrelations on a one-dimensional cluster
state.  A final section summarizes our conclusions.

\section{Graph states}
\label{sec:states}

A graph is a set of $n$ nodes and a set of edges connecting them. The
neighborhood $\nghb(j)$ of a node $j$ is the set of nodes that are
connected to it. Given a particular graph, we can associate a qubit with
each node and define the corresponding graph state of the qubits in the
following way.  Let $X$, $Y$, and $Z$ denote the Pauli matrices $\sigma_x$, $\sigma_y$, and $\sigma_z$, and adopt the shorthand of writing tensor products of Pauli matrices as products of Pauli matrices indexed by position, i.e. $X\otimes I\otimes Y = X_1Y_3$.  The graph
state $\ket{\psi}$ on $n$ qubits is the simultaneous $+1$ eigenstate of
the (commuting) operators
\begin{equation}
G_j=X_j\prod_{k \in\nghb(j)}Z_k\;,\quad j=1,\ldots,n.
\label{eq:gn}
\end{equation}
The operators $G_j$ constitute an independent set of generators of the
stabilizer group of $\ket{\psi}$.  Any graph state can be constructed by
preparing each qubit in the eigenstate of spin up in the $x$ direction and
then applying a controlled-phase gate between each pair of qubits that is
connected by an edge in the graph.  The order in which the
controlled-phase gates are implemented is unimportant since they all
commute.

The structure of graph states makes them good candidates for the study
of nonlocality.  For a connected graph (of at least two nodes), all single-qubit measurements
yield random values, yet these values are correlated in such a way that
certain products of them give deterministic results.  If $M$ represents
an $n$-fold tensor product of the Pauli matrices, $I$, $X$, $Y$, and
$Z$, then the result of measuring $M$ on the $n$-qubit graph state
$\ket{\psi}$ is determined by which of following three cases applies to
$M$ (see Ref.~\cite{nielsenchuang}):
\begin{enumerate}
\item[(i)]$M$ is an element of the stabilizer group, i.e., $M=G_1^{a_1}\cdots
G_n^{a_n}$ is a product of the generators $G_j$ for some $a_j=0,1$, in
which case a measurement of $M$ obviously gives outcome~$+1$.
\item[(ii)]$-M$ is an element of the stabilizer group, i.e., $-M=G_1^{a_1}\cdots
G_n^{a_n}$ is a product of the generators $G_j$ for some $a_j=0,1$, in
which case a measurement of $M$ obviously gives outcome~$-1$.
\item[(iii)]$\pm M$ is not an element of the stabilizer, i.e., $M$ is not a
product of the generators up to a multiplicative factor $\pm 1$, in
which case a measurement of $M$ gives outcomes $+1$ and $-1$ with equal
probability.
\end{enumerate}
The minus sign in case~(ii) comes from the fact that products of
generators can introduce at each site terms such as
$ZXZ=-X$ or $ZX=iY$, with $i$'s from pairs of sites
multiplying to give a $-1$. These terms lead to GHZ-like paradoxes for
the graph state, implying that communication between the parties is
required to model the correlations classically.

\section{Communication-assisted LHV model for graph states}
\label{sec:model}

\subsection{Description of the model}
\label{subsec:model}

Our model uses $n$ binary random variables, $z_1,\ldots,z_n$, each taking
on values $\pm1$ with equal probability.  These hidden variables can be
thought of as values for the $z$ spin components of the $n$ qubits. For
the corresponding values of the $x$ and $y$ spin components, we define the
quantities
\begin{subequations}
\label{eq:lhv}
\begin{eqnarray}
x_j&=&\prod_{k\in\nghb(j)}z_{k}\;,
\label{eq:lhvx}\\
y_j&=&z_j\prod_{k\in\nghb(j)}z_{k}
\label{eq:lhvy}\;.
\end{eqnarray}
\end{subequations}
The values $x_j$ are suggested by the $+1$ values associated with
the generators $G_j$, i.e.,
\begin{equation}
x_j\prod_{k\in\nghb(j)}z_k=+1\;,
\end{equation}
in analogy to Eq.~(\ref{eq:gn}).
The values
\begin{equation}\label{yhv}
y_j=x_jz_j
\end{equation}
are suggested by the analogous relations $Y_j=iX_jZ_j$ for Pauli matrices.

We assume now that each party is given a measurement $M_j$ to perform,
chosen from $I$ (no measurement), $X$, $Y$, and $Z$.  After the
measurement, there is a round of communication between neighboring sites,
and then each party outputs a value $+1$ or $-1$ as the result of the
measurement.  When no measurement is performed at a site, the output can
be regarded as $+1$.

During the round of communication, site~$j$ sends a bit $c_j$ to each site
$k\in\nghb(j)$, where $c_j=0$ if $M_j=I,Z$ and $c_j=1$ if
$M_j=X,Y$.  The value $v_j$ output at site~$j$ is determined by the
hidden variable for the observable measured at that site and by the
quantity
\begin{equation}
t_j=\sum_{k\in\nghb(j)}c_k\mod4\;,
\label{eq:nj}
\end{equation}
which is computed from the bits sent to site~$j$ from neighboring sites
and which is equal to the number of neighboring sites that make an $X$
or $Y$ measurement modulo~4.  The output $v_j$ is determined by rules that
decide whether to flip the sign of the hidden variable associated with
the measurement at site~$j$:
\begin{enumerate}
\item If $M_j=I$, $v_j=1$.
\item If $M_j=Z$, $v_j=z_j$.
\item If $M_j=X$, $v_j=\left\{\begin{array}{rl}x_j &\mathrm{if} \quad t_j=0,1,\\
-x_j &\mathrm{if} \quad t_j=2,3.\end{array}\right.$
\item If $M_j=Y$, $v_j=\left\{\begin{array}{rl}y_j &\mathrm{if} \quad t_j=1,2,\\
-y_j &\mathrm{if} \quad t_j=0,3.\end{array}\right.$
\end{enumerate}
This protocol reproduces the quantum predictions for any global Pauli
measurement on graph states, as we show in the next subsection. In
other words, if we take the product of the outputs from all the sites,
the result is the same as the quantum prediction for a measurement of
the operator $M=\bigotimes_{j=1}^n M_j$.  The number of bits
communicated in this protocol is twice the number of edges in the
graph.

Variants of rules~3 and~4 also give the correct predictions for global
Pauli measurements; for example, these rules can be modified so that
the sign flip occurs under the same circumstances for both $X$ and $Y$
measurements.  We note, however, that neither the rules given above nor
these modified rules are guaranteed to reproduce all of the
correlations predicted by quantum mechanics on subsets of the Pauli
operators measured. We take up the question of these subcorrelations in
Sec.~\ref{sec:failure}.

\subsection{Proof that the model works}
\label{subsec:proof}

The proof that our model yields the correct global quantum
predictions proceeds in two stages.  We first introduce a simple
related model that involves no classical communication and show that
this simple model makes the correct quantum predictions in cases~(i)
and~(iii) above, but not in case~(ii).  We then show that the
communication-assisted model makes correct global predictions in all
three cases.

The simple no-communication model has each party output the hidden
variable $1$, $x_j$, $y_j$, or $z_j$ associated with the measurement
made at its site.  The communication-assisted model is derived from
this no-communication model by sometimes flipping the sign of the
outcome at a site where $X$ or $Y$ is measured, i.e., by outputting
$-x_j$ instead of $x_j$ or $-y_j$ instead of $y_j$.  The decision to
flip a sign is determined by the number of $X$ and $Y$ measurements
made at neighboring sites, in accordance with the conditions in rules~3
and~4.

For any tensor product of Pauli operators, $M=\bigotimes_{j=1}^n M_j$,
it is useful to introduce a corresponding $n$-tuple $m$ of the same
form as $M$, but with the tensor product of Pauli operators replaced by
an $n$-tuple of the corresponding hidden variables, $1$, $x_j$, $y_j$,
and $z_j$.  In the no-communication model, the elements of the
$n$-tuple are the outcomes of the measurements $M_j$.  The $n$-tuples
$m$ form an abelian group of order $4^n$, with multiplication defined
bitwise.

The hidden variables $x_j$, $y_j$, $z_j$ satisfy a commutative algebra,
similar to the Pauli algebra, with $x_j^2=y_j^2=z_j^2=1$ and
$y_j=x_jz_j$, as in Eq.~\eqref{yhv}. The noteworthy differences from
the Pauli algebra are the commutativity and the absence of an $i$ in
Eq.~\eqref{yhv}. As a consequence, when a measurement has the form
$M=\pm G_1^{a_1}\cdots G_n^{a_n}$, the product of all parties' outputs
in the no-communication model always equals $+1$.  Thus it is clear
that the no-communication model gets the correct result in case~(i)
above, but not in case~(ii).

We show now that the no-communication model is also correct in
case~(iii).  For this purpose, note that the $n$-tuples $g_j$
associated with the stabilizer generators $G_j$ generate a subgroup of
order $2^n$, which contains the $n$-tuples associated with all
Pauli products $M$ such that $\pm M$ is in the stabilizer.  This subgroup
defines $2^n$ cosets which, except for the subgroup itself, necessarily
contain $n$-tuples associated with Pauli products from case~(iii).
Thus we need to show that the no-communication model predicts a random
outcome for all cosets except the subgroup itself.  We note that two
$n$-tuples in the same coset predict the same outcome, thus allowing us
to restrict attention to a single element in each coset.  Elements of
the form $(z_1^{a_1},\ldots, z_n^{a_n})$, with $a_j=0,1$, clearly predict
a random overall outcome, except when $a_j=0$ for all~$j$ (i.e., the
identity $n$-tuple).  Moreover, these $2^n$ $n$-tuples each belong to a
different coset, since they make up a subgroup of their own that
contains none of the elements of the subgroup generated by the $g_j$,
except the identity.  Thus we recover the correct predictions for
case~(iii).

The next step in the proof is to show that the communication-assisted
LHV model recovers the correct predictions for a measurement of $M$ in
all three cases.  If $M$ is as in~(iii), then the result predicted by
the no-communication model is random, and flipping an outcome at any
site does not affect this.  Thus the communication model works when $M$
is as in~(iii).  To show that the model also works when $M$ is as
in~(i) or~(ii), we proceed by induction.  The model works when $M$ is
any one of the generators $G_j$.  We consider a Pauli product
$M=\bigotimes_{l=1}^n M_l$ that is a product, up to a factor $\pm1$, of
generators $G_k$ with $k<j$.  With this assumption, it is clear that
$M_j$ is either $I$ or $Z$.  Our inductive procedure is to show that
if the model correctly predicts the overall correlation for $M$, then
it also reproduces the overall correlation for $M'=\pm MG_j$.

Consider the outcome for a measurement of $M'$, as predicted by quantum
mechanics.  We express $M'$ in terms of $M=\bigotimes_{l=1}^n M_l$ and
the generator $G_j=X_j\prod_{k\in\nghb(j)}Z_k$.  Upon multiplying
$M$ with $G_j$, we obtain the following: (a)~at each $k\in\nghb(j)$ for
which $M_k=I$, the product $IZ=Z$ gives $M'_k=Z$; (b)~at each
$k\in\nghb(j)$ for which $M_k=Z$, the product $ZZ=I$ gives
$M'_k=I$; (c)~at each $k\in\nghb(j)$ for which $M_k=X$---we let $q$
denote the number of such sites---the product $XZ=-iY$ gives
$M'_k=Y$ and introduces a factor of $-i$; (d)~at each $k\in\nghb(j)$
for which $M_k=Y$---we let $r$ denote the number of such sites---the
product $YZ=iX$ gives $M'_k=X$ and introduces a factor of $+i$.
Overall we thus obtain a factor $(-i)^qi^r=(-i)^{q+r}(-1)^r$. Now
consider site~$j$: if $q+r$ is even, $M_j=I$, and we are left with
$M'_j=X$ and no additional factors of $i$; if $q+r$ is odd,
$M_j=Z$, and we are left with $M'_j=Y$ and an additional factor of
$i$.  There are thus four possibilities: if $q+r=0,1\!\!\mod4$, then
$M'=(-1)^rMG_j$, and if $q+r=2,3\!\!\mod4$, then $M'=(-1)^{r+1}MG_j$.
It follows that in the case $q+r=0\!\!\mod4$, the quantum prediction
for measurement of the operator $M'$ is equal to the quantum prediction
for a measurement of $M$ multiplied by $(-1)^r$, and similarly for the
other cases.

Now we consider the prediction given for a measurement of $M'$ by our
communication-assisted LHV model, assuming that the correct prediction
is returned for $M$.  The value returned for a measurement of $M'$ is
equal to the value returned for a measurement of $M$, multiplied by the
value returned for a measurement of $G_j$, which is 1, and by a $-1$
for each site that changes its sign-flip decision.  A review of the
immediately preceding paragraph shows that the only site that changes
the $c$-bit sent to neighboring sites is site~$j$, which changes its
$c$-bit from $c_j=0$ to $c_j=1$.  This means that at neighboring sites
$k\in\nghb(j)$, the quantity $t_k$ of Eq.~(\ref{eq:nj}) increases by 1.
At neighboring sites $k\in\nghb(j)$ for which $M_k=X$, the $X$
becomes a $Y$ in $M'$, with $t_k$ increased by 1, so according to
rules~3 and~4, there is no change in the sign-flip decision.  At
neighboring sites $k\in\nghb(j)$ for which $M_k=Y$, the $Y$ becomes
a $X$ in $M'$, with $t_k$ increased by 1, so according to rules~3
and~4, site $k$ changes its sign-flip decision.  The result of these
changes is an overall factor of $(-1)^r$.  The final contribution comes
from site~$j$, for which $t_j=q+r \mod 4$, and which changes from
$M_j=I$ to $M'_j=X$ if $q+r$ is even, and from $M_j=Z$ to
$M'_j=Y$ if $q+r$ is odd.  According to the rules, the effect of
these changes is to introduce an additional sign flip if and only if
$q+r=2,3\!\!\mod4$, which is just what is required to return the
quantum predictions.

\section{Graph-state submeasurements}
\label{sec:failure}

Having shown that our communication-assisted LHV model agrees with quantum
mechanics for global correlations, we now consider the question of
submeasurements.  A submeasurement of a global Pauli product~$M$ is a
Pauli product $\tilde M$ such that the non-identity elements of $\tilde M$
all appear in $M$, i.e., $\tilde M_j=M_j$ or $\tilde M_j=I$ for all~$j$.
LHV models implicitly predict the result of measuring such a subset of the
Pauli operators of a global measurement, the measurement of an identity
operator being simply the omission of the corresponding local hidden
variable.  A proper communication-assisted LHV model for graph states
should not only reproduce the predictions of quantum mechanics for global
measurements but also for all possible submeasurements.

It can be shown that our model satisfies this condition for some, but not
all, graph states.  Determining the graphs for which it works, a class
including complete bipartite graphs (a case encompassing the star graphs
of GHZ states) and the symmetric difference of two complete
graphs~\cite{diestel}, requires the introduction of techniques otherwise
unused in this paper, and, as such, we reserve its exposition for another
time~\cite{adjacency}.  Instead, we focus here on understanding the
properties that limit our model's effectiveness.  The following two
subsections show that protocols with fixed communication distance or with
symmetric communication and decision protocols generally do not reproduce
all subcorrelations on all graphs.  The final subsection further explores
the symmetry of site invariance by considering it in the context of
one-dimensional cluster states.

\subsection{Non-nearest-neighbor communication protocols}
\label{subsec:triangle}

In a communication protocol with communication distance $d$, nodes $j$ and
$k$ can signal to each other if there exists within the graph a path from
$j$ to $k$ that traverses $d$ or fewer edges~\cite{diestel}.  Put another
way, this is the statement that information can only be transmitted along
edges and that the number of successive edges through which some piece of
information can be sent is at most $d$.  In this section we prove, via
contradiction, that no communication-assisted LHV model for which the
communication distance satisfies
\begin{equation}
  d\leq4\left\lfloor \frac{n}{24}-\frac{1}{2}\right\rfloor+1
\end{equation}
correctly reproduces the predictions of quantum mechanics for all
submeasurements on all graph states of $n$ qubits.

The proof relies on an infinite class of graph states for which a set of
five global measurements can be chosen that are not locally
distinguishable.  Each of these global measurements includes a
submeasurement that can be written in terms of stabilizer elements and is
thus certain.  The output of each qubit, however, must be such that the
correct values are obtained for all submeasurements that are consistent
with its observable surroundings.  This requirement, for the particular
states and measurements chosen, yields a contradiction.

To begin, consider the graph state corresponding to an $n$-node ring where
$n=12f$ and $f$ is an odd positive integer.  Let the qubits be numbered
sequentially, starting with $1$ at an arbitrary point on the ring and
moving clockwise along it.  Additionally, define the following subsets of
the $n$ labels:
\begin{subequations}
\begin{align}
  \mathcal{V}&=\{4f,8f,12f\}\;,\\
  \mathcal{M}&=\{2f,6f,10f\}\;,\\
  \mathcal{Y}&=\{j|j\equiv1\text{ mod }2\}\;,\\
  \mathcal{L}&=\{j|j\not\in\mathcal{V},\mathcal{M}\text{ and }j\equiv2\text{ mod }4\}\;,\\
  \mathcal{R}&=\{j|j\not\in\mathcal{V},\mathcal{M}\text{ and }j\equiv0\text{ mod }4\}\;,\\
  \mathcal{S}_k&=\{j|2f(k-1)<j<2f k\}\;.
\end{align}
\end{subequations}
For our purposes, it is useful to think of the ring as arranged in an
equilateral triangle with vertices specified by the subset $\mathcal{V}$
(see figure \ref{fig:triangle}). The midpoints of the legs of the triangle
are then given by the subset $\mathcal{M}$, and the segments between
adjacent vertices and midpoints are given by the $\mathcal{S}_j$'s.  We
use the notation $\mathcal{S}_{j,k}$ as shorthand for
$\mathcal{S}_j\cup\mathcal{S}_k$, and $\mathcal{A}\backslash\mathcal{B}$
is used to denote the set consisting of the elements of $\mathcal{A}$ that
are not in $\mathcal{B}$.

Now consider global measurements of the form
\begin{equation}
M_j =
  \begin{cases}
    X\text{ or }Y & \text{if } j\in\mathcal{V},\\
    Y & \text{if } j\in\mathcal{Y},\\
    X & \text{otherwise,}
  \end{cases}\label{eq:globalmeas}
\end{equation}
such that the number of vertices measuring $Y$ is not one.  These global
measurements include the following submeasurements, for which quantum mechanics
predicts an outcome with certainty.\\

\begin{figure}
\[
 \Qcircuit[2.2em]@R=1.5em @C=0em{
 & & & & \node[12]{\raisebox{.11em}{$\scriptstyle{X/Y}$}} \link{1}{-1} \link{1}{1} & & & & \\
 & & & \node[11]{Y} \link{1}{-1} & & \node[1]{Y} \link{1}{1} & & & \\
 & & \node[10]{X} \link{1}{-1} & & & & \node[2]{X} \link{1}{1} & & \\
 & \node[9]{Y} \link{1}{-1} & & & & & & \node[3]{Y} \link{1}{1} & \\
 \node[8]{\raisebox{.11em}{$\scriptstyle{X/Y}$}} \link{0}{2} & & \node[7]{Y} \link{0}{2} & & \node[6]{X} \link{0}{2} & & \node[5]{Y} \link{0}{2} & & \node[4]{\raisebox{.11em}{$\scriptstyle{X/Y}$}}}
\]
\caption{Example demonstrating that any communication-assisted LHV model with communication distance $d \leq 1$ fails to reproduce some submeasurements for the ring with $n = 12$ nodes. Five of the global measurements shown on the \vspace{.4em}ring,
\vspace{.4em}\centerline{${\color{gray}Y_1}X_2{\color{gray}Y_3}X_4{\color{gray}Y_5}X_6{\color{gray}Y_7}X_8{\color{gray}Y_9}X_{10}{\color{gray}Y_{11}}X_{12}$\;,}
\vspace{.4em}\centerline{$Y_1X_2Y_3{\color{gray}Y_4}Y_5X_6Y_7{\color{gray}Y_8}Y_9X_{10}Y_{11}{\color{gray}Y_{12}}$\;,}
\vspace{.4em}\centerline{$Y_1{\color{gray}X_2}Y_3Y_4{\color{gray}Y_5}X_6{\color{gray}Y_7}X_8{\color{gray}Y_9}X_{10}{\color{gray}Y_{11}}Y_{12}$\;,}
\vspace{.4em}\centerline{${\color{gray}Y_1}X_2{\color{gray}Y_3}Y_4Y_5{\color{gray}X_6}Y_7Y_8{\color{gray}Y_9}X_{10}{\color{gray}Y_{11}}X_{12}$\;,}
\vspace{.4em}\centerline{${\color{gray}Y_1}X_2{\color{gray}Y_3}X_4{\color{gray}Y_5}X_6{\color{gray}Y_7}Y_8Y_9{\color{gray}X_{10}}Y_{11}Y_{12}$\;,}
contain submeasurements (shown in black) useful for showing a contradiction.  These
submeasurements imply the following constraints on a nearest-neighbor communication \vspace{.2em}model,
\vspace{.4em}\centerline{\hspace{3.2em}$x_2^{}x_4^{}x_6^{}x_8^{}x_{10}^{}x_{12}^{}=1$\;,\hspace{1em}}
\vspace{.4em}\centerline{$y_1^Yx_2^{}y_3^Yy_5^Yx_6^{}y_7^Yy_9^Yx_{10}^{}y_{11}^Y=-1$\;,\hspace{1em}}
\vspace{.4em}\centerline{\hspace{1.9em}$y_1^Yy_3^Yy_4^{}x_6^{}x_8^{}x_{10}^{}y_{12}^{}=1$\;,\hspace{1em}}
\vspace{.4em}\centerline{\hspace{1.9em}$x_2^{}y_4^{}y_5^Yy_7^Yy_8^{}x_{10}^{}x_{12}^{}=1$\;,\hspace{1em}}
\vspace{.4em}\centerline{\hspace{2.5em}$x_2^{}x_4^{}x_6^{}y_8^{}y_9^Yy_{11}^Yy_{12}^{}=1\;,$\hspace{1em}}
which when multiplied together yield a contradiction.
\label{fig:triangle}}
\end{figure}

\begin{subequations}\label{eq:allcertain}
\noindent For $M_{4f}$, $M_{8f}=X$, $M_{12f}=X$,
\begin{align} \label{eq:XXX}
X_{2f}X_{4f}X_{6f}X_{8f}X_{10f}X_{12f}\prod_{j\in\mathcal{L}\cup\mathcal{R}} X_{j} = \prod_{j=1}^{6f} G_{2j}\;,
\end{align}
implying a measurement outcome of $+1$.\\

\noindent For $M_{4f}=Y$, $M_{8f}=Y$, $M_{12f}=Y$,
\begin{align} \label{eq:YYY}
\begin{split}
X_{2f}&X_{6f}X_{10f}\prod_{j\in\mathcal{Y}} Y_{j} \prod_{k\in\mathcal{L}} X_{k} \hspace{8em} \\
&= \prod_{j=1}^{3f} \Bigl(-G_{4j-3}G_{4j-2}G_{4j-1}\Bigr)\;,
\end{split}
\end{align}
implying a measurement outcome of $(-1)^{3f} = -1$.\\

\noindent For $M_{4f}=Y$, $M_{8f}=X$, $M_{12f}=Y$,
\begin{align} \label{eq:YXY}
\begin{split}
Y_{4f}X_{6f}X_{8f}X_{10f}Y_{12f}\prod_{j\in\mathcal{Y}\cap\mathcal{S}_{1,2}} Y_j
\prod_{k\in\mathcal{R}\cup(\mathcal{L}\backslash\mathcal{S}_{1,2})} X_k \\
= G_1 \prod_{j=1}^{f-1} \Bigl(-G_{4j-1}G_{4j}G_{4j+1}\Bigr) G_{4f-1} \prod_{k=2f}^{6f} G_{2k}\;,
\end{split}
\end{align}
implying a measurement outcome of $(-1)^{f-1} = +1$.\\

\noindent Cyclic permutation of this last measurement yields two more with $+1$ outcomes.\\

\noindent For $M_{4f}=Y$, $M_{8f}=Y$, $M_{12f}=X$, we have
\begin{align} \label{eq:YYX}
X_{2f}Y_{4f}Y_{8f}X_{10f}X_{12f}\prod_{j\in\mathcal{Y}\cap\mathcal{S}_{3,4}} Y_j
\prod_{k\in\mathcal{R}\cup(\mathcal{L}\backslash\mathcal{S}_{3,4})} X_k\;,
\end{align}\\

\noindent and for $M_{4f}=X$, $M_{8f}=Y$, $M_{12f}=Y$, we have
\begin{align} \label{eq:XYY}
X_{2f}X_{4f}X_{6f}Y_{8f}Y_{12f}\prod_{j\in\mathcal{Y}\cap\mathcal{S}_{5,6}} Y_j
\prod_{k\in\mathcal{R}\cup(\mathcal{L}\backslash\mathcal{S}_{5,6})} X_k\;.
\end{align}

\end{subequations}

Now assume there exists a distance $d=2f-1$ communication-assisted LHV
model that correctly replicates the predictions of quantum mechanics for
all Pauli measurements on $n$ qubits.  The output of such a model can be
fully described in terms of single-qubit hidden variables whose value
depends both on the qubit in question and on the measurements made by
other qubits within its communication range.  We write these hidden
variables in the form ${\sigma_{\!j}}^\alpha$ where $j$ is the qubit being
measured, $\sigma$ is the hidden variable corresponding to the Pauli
operator measured upon it, and $\alpha$ indicates the measurements made on
qubits within its communication range.  The global measurements utilized
for Eqs.~(\ref{eq:allcertain}) have the virtue that each qubit's
communication range includes at most one other qubit whose measurement is
changeable, and that is the qubit at the nearest vertex.  Thus, in
comparisons between them, the measurement performed on, at most, a single
qubit need be included in $\alpha$.  Moreover, the qubits at the center of
each side of the triangle cannot see the changes at the vertices.
Consequently, the constraints implied by Eqs.~(\ref{eq:allcertain}) on a
hidden variable model with communication range $d$ can be expressed as
follows:

\begin{subequations}
\begin{align}
1=&x_{2f}x_{4f}x_{6f}x_{8f}x_{10f}x_{12f}\prod_{j\in\mathcal{L}\cup\mathcal{R}} x_{j}^X\;, \label{eq:xxx}\\
-1=&x_{2f}x_{6f}x_{10f} \prod_{j\in\mathcal{Y}} y_{j}^Y \prod_{k\in\mathcal{L}} x_{k}^Y\;, \\
\begin{split}
1=&y_{4f}x_{6f}x_{8f}x_{10f}y_{12f}
\prod_{j\in\mathcal{Y}\cap\mathcal{S}_{1,2}} y_j^Y \\
&\hspace{2em} \prod_{k\in(\mathcal{L}\cup\mathcal{R})\cap\mathcal{S}_{4,5}} x_k^X
\prod_{l\in(\mathcal{L}\cap\mathcal{S}_{3,6})\cup(\mathcal{R}\backslash\mathcal{S}_{4,5})} x_l^Y\;,
\end{split}\\
\begin{split}
1=&x_{2f}y_{4f}y_{8f}x_{10f}x_{12f}
\prod_{j\in\mathcal{Y}\cap\mathcal{S}_{3,4}} y_j^Y \\
&\hspace{2em}\prod_{k\in(\mathcal{L}\cup\mathcal{R})\cap\mathcal{S}_{6,1}} x_k^X
\prod_{l\in(\mathcal{L}\cap\mathcal{S}_{5,2})\cup(\mathcal{R}\backslash\mathcal{S}_{6,1})} x_l^Y\;,
\end{split}\\
\begin{split}
1=&x_{2f}x_{4f}x_{6f}y_{8f}y_{12f}
\prod_{j\in\mathcal{Y}\cap\mathcal{S}_{5,6}} y_j^Y \\
&\hspace{2em}\prod_{k\in(\mathcal{L}\cup\mathcal{R})\cap\mathcal{S}_{2,3}} x_k^X
\prod_{l\in(\mathcal{L}\cap\mathcal{S}_{1,4})\cup(\mathcal{R}\backslash\mathcal{S}_{2,3})} x_l^Y\;.
\end{split}
\end{align}
\end{subequations}

Using the identity
$\mathcal{A}=\mathcal{A}\cap(\mathcal{S}_1\cup\mathcal{S}_2\cup\mathcal{S}_3\cup\mathcal{S}_4\cup\mathcal{S}_5\cup\mathcal{S}_6)$
for $\mathcal{A}=\mathcal{Y}\text{, }\mathcal{L}\text{, or }\mathcal{R}$
and the fact that all variables square to $1$, it can be shown that the
right-hand side of Eq.~(\ref{eq:xxx}) is equal to the product of the
right-hand sides of the other four equations.  Thus, we have the
contradiction $1=-1$, showing that no distance-$d$ communication-assisted
LHV model reproduces the predictions of quantum mechanics in this
instance.

For other values of $n \neq 12f$, with $f$ odd, an identical contradiction
applies to a graph consisting of $r=(n-12)\text{ mod }24$ unconnected
nodes and a ring of size $n-r$. It is also possible to adapt our example to two-dimensional cluster states. One can show, for example, that for a $(3f+3)\times (3f+3)$ cluster state, with $f$ odd, a communication distance of at least $2f$ is required.

\subsection{Site-invariant communication protocols}
\label{subsec:invariant}

Both the numbering and the arrangement of nodes in a graph are arbitrary,
so it seems reasonable to suppose that a communication protocol should be
insensitive to these things.  We refer to this property as site invariance
and define it formally as follows.  Given a graph $\mathcal{G}$, each of
whose nodes has been assigned a measurement, a permutation that leaves the
graph invariant is one that interchanges nodes and their measurements,
letting edges move with the nodes, such that the new graph $\mathcal{G}'$
is identical to $\mathcal{G}$ in the sense that they could be placed on
top of each other with all nodes, measurements, and edges overlapping.  A
site-invariant protocol is one for which nodes in identical situations, as
defined by permutations that leave the graph invariant, make the same
sign-flipping decision.  Surprisingly, we find this trait to be at odds
with the modeling of submeasurements.

\begin{figure}
\[
\Qcircuit[2.2em]@R=0em @C=0em{
\node[1]{Y} & & \node[2]{Y} \link{0}{-2} & & \node[3]{Y} \link{0}{-2} \\ \\
\nodeDown[6]{Y} \link{-2}{0} & & \nodeDown[5]{Y} \link{0}{-2} \link{-2}{0} & &
\nodeDown[4]{Y} \link{0}{-2} \link{-2}{0}}
\]
\caption{Example demonstrating that any communication-assisted LHV model
based on the hidden variables of Eq.~(\ref{eq:lhv}) and assisted by a
site-invariant communication protocol fails to reproduce some
submeasurements. The global measurement $M=Y_1Y_2 Y_3Y_4Y_5Y_6$ has a
random outcome, but contains a submeasurement $\tilde M=Y_1Y_2Y_3
I_4Y_5I_6$ such that $-\tilde{M}$ is an element of the stabilizer group.
This means that an overall sign flip is required to correct the $+1$
prediction of the hidden variables for a measurement of $\tilde M$. The
two qubits measuring $Y$ at nodes $1$ and $3$ are in symmetric situations,
as are the qubits at nodes $2$ and $5$. Thus, under a site-invariant
protocol, $1$ and $3$ must make the same sign-flipping decision, as must
$2$ and $5$. For each pair, the sign-flipping decisions cancel one
another, producing no overall sign flip and thus giving an incorrect
result of $+1$ for the measurement of $\tilde M$. \label{fig:cluster}}
\end{figure}
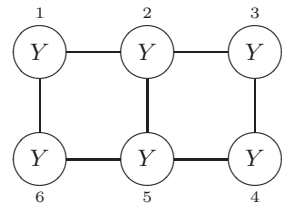

We demonstrate the limitations imposed by site invariance using the
example of a $2\times3$ cluster state, which is depicted in
Fig.~\ref{fig:cluster}.  The two relevant measurements for this example
are $M=Y_1Y_2Y_3Y_4Y_5Y_6$, which has a random outcome, and $\tilde
M=Y_1Y_2Y_3 I_4Y_5I_6$, which has the certain outcome $-1$.  When either
of these is considered as a global measurement, our model yields the
correct prediction, as we have already shown in general, but when the
second is considered as a submeasurement of the first, the model fails. In
this second case, rules~1--4 say that the two qubits measuring $Y$ at
nodes $2$ and $5$ should introduce a sign flip, but the two qubits
measuring $Y$ at nodes $1$ and $3$ should not.  The result is no overall
sign flip and an outcome $+1$, showing that the model gets the
submeasurement outcome wrong.  In contrast, when $\tilde M$ is considered
as a global measurement, rules~1--4 dictate a sign flip for qubit $2$, but
no other qubit, thus giving the correct, certain outcome $-1$.  The same
measurement $\tilde M$ can lead to different sign-flipping decisions in
the two situations because the nearest-neighbor environments of the qubits
differ depending on whether a submeasurement or a global measurement is
under consideration.  As is shown in Figure~\ref{fig:cluster}, the
counterexample is not limited to the communication model used in this
paper.  In fact, any site-invariant protocol based on our hidden variables
yields an incorrect result for the submeasurement $\tilde M$.

This example can easily be generalized by adding $p$ rows and $q$ columns
to opposite sides of the $2 \times 3$ cluster state. Doing this results in
a class of $(2+2p) \times (3+2q)$ cluster states for which LHV models
based on the hidden variables of Eq.~(\ref{eq:lhv}) and assisted by a
site-invariant communication protocol fail for some submeasurements.

\subsection{Site-invariant model for 1-D cluster states}
\label{subsec:cluster}

In Sec.~\ref{subsec:triangle} it was shown that there exist graph states
of size $n$ for which any communication-assisted LHV model must involve
communication over a distance at least $n/6$ if it reproduces all
subcorrelations.  Note that this result applies to all models, whether
site-invariant or not, and whatever the structure of the LHVs. In
Sec.~\ref{subsec:invariant} it was shown that for certain graph states, no
model based on the hidden variables of Eq.~(\ref{eq:lhv}) and assisted by
a site-invariant communication protocol is capable of reproducing all
subcorrelations.  This result holds even if the model allows
unlimited-distance communication.

It is intriguing that both of these results apply to two-dimensional
cluster states, since two-dimensional cluster states, along with
single-qubit measurements, are universal for quantum
computation.\footnote{Admittedly, our models are only concerned with
measurements of Pauli operators, which are not universal for computation
due to the Gottesman-Knill theorem \cite{gk}.} It is therefore reasonable
to ask whether the same results hold for one-dimensional cluster states
(linear chains), which are not universal for computation \cite{universal}.
In this subsection we show that linear chains do permit successful
site-invariant protocols. The protocol we describe involves communication
over a distance equal to the number of edges in the one-dimensional
cluster state (i.e, unlimited communication).  At present it is unknown
whether the subcorrelations of a linear chain could be reproduced by a
protocol with limited-distance communication.

The fact that unlimited communication is allowed is in the same spirit as our
counterexample of Fig.~\ref{fig:cluster}, where communication spans the
entire graph and the only restriction is site invariance.  The key
simplification in the case of one-dimensional cluster states is that all
qubits, except those at the ends of the chain, have exactly two neighbors.
As a consequence, the form of stabilizer elements whose hidden-variable
result from Eq.~(\ref{eq:lhv}) requires correction is constrained so that the correction can
be effected by a site-invariant protocol.

For an $n$-qubit chain, the $n$ stabilizer generators are given by
$G_1=X_1Z_2$, $G_j=Z_{j-1}X_jZ_{j+1}$ for $j=2,\ldots,n-1$, and
$G_n=Z_{n-1}X_n$.  Any stabilizer element is a product of generators.  An
arbitrary product of generators can be decomposed into a product of terms
each of which is a product of successive generators.  We call these terms
{\em primitive stabilizers\/} or just {\em primitives}.  The primitive
stabilizers are separated by the omission of one or more generators in the
product of generators.  An example of a stabilizer element for $n=10$
qubits is $G_1G_2G_3G_5G_6G_9=-Y_1X_2Y_3I_4Y_5Y_6Z_7Z_8X_9Z_{10}$.
The primitives in this example are $G_1G_2G_3$, $G_5G_6$, and
$G_9$.

Associated with each primitive is a Pauli product (with the sign omitted)
for the qubits corresponding to the generators in the primitive.  We call
these Pauli products {\em words}.  For the 10-qubit example above, the
words are $Y_1X_2Y_3$, $Y_5Y_6$, and $X_9$.  At each end of a word, there is an $I$ if
one generator is omitted and a $Z$ if two or more generators are omitted.
We can make these word boundaries apply even at the end of the linear
chain by embedding our cluster state in an infinite linear chain.  The
generators for the qubits to the left of $j=1$ and to the right of $j=n$
are always omitted, and we redefine $G_1=Z_0X_1Z_2$ and
$G_n=Z_{n-1}X_nZ_{n+1}$.

If a word is bounded by an $I$, there must be another word immediately on
the other side of the $I$.  A {\em sentence\/} is a Pauli product
consisting of a set of words separated by singleton $I$s and bracketed by
$Z$s at both ends.  Words are not stabilizer elements, but sentences are.
The example above contains two sentences, $Z_0Y_1X_2Y_3I_4Y_5Y_6Z_7$ (including the zeroth
qubit) and $Z_8X_9Z_{10}$.  The $Z$ bookends on a sentence separate it from other,
nonoverlapping sentences in the same overall stabilizer element.  Between
the $Z$s in successive sentences, there can be an arbitrary number of
$I$s. Any stabilizer element is a product of nonoverlapping sentences.

We can list the entire set of words by considering all possible
primitives:
\begin{subequations}
\label{eq:primitives}
\begin{align}
&\hspace{-12pt}\parbox{20em}{$X$ for a primitive with one Pauli operator;}\\
&\hspace{-12pt}\parbox{20em}{$Y\otimes Y$ for a primitive with two Pauli operators;}\\
&\hspace{-12pt}\parbox{20em}{$Y\otimes X^{\otimes(j-2)}\otimes Y$ for a primitive with $j\ge3$ Pauli operators.}\label{eq:primc}
\end{align}
\end{subequations}
For stabilizer elements, $I$s occur only between sentences or as
singletons between words, $X$s and $Y$s occur only in words, and $Z$s
occur only as the boundaries of sentences.

Recall that the goal of the communication protocol is to introduce a sign
flip into the product of hidden-variable entries for those Pauli products
that are the negative of a stabilizer element.  The only words that
introduce a minus sign into the corresponding product of generators are
those of the form $Y\otimes X^{\otimes(j-2)}\otimes Y$ with $j$ odd.  Thus a candidate for
a site-invariant communication protocol is the following.
\begin{enumerate}
\item Each site at which an $X$ or a $Z$ is measured broadcasts the measurement performed upon it.
\item Each site that measures $X$ determines if it is the middle (implying an odd number of $X$s) qubit
in a word of the form~(\ref{eq:primc}) in a submeasurement sentence, and if so, flips
its hidden-variable entry, i.e., changes $x$ to $-x$.
\end{enumerate}
This clearly gets any stabilizer right and thus all global correlations
right.

The only question remaining is whether this protocol works for
subcorrelations.  We answer this question by showing the following: {\em
two sentences, $S_1$ and $S_2$, that are submeasurements of the same
global measurement, generally not a stabilizer element, must be identical
on the region where they overlap, except possibly at bracketing $Z$s.}
This property implies that $S_1$ and $S_2$ have exactly the same words in
the region of overlap.  Thus, for any pair of submeasurements of the global
measurement, a sign flip arising from a word of the form~(\ref{eq:primc})
in the overlap region is common to both submeasurements.  Since both the
word and the sign flip occur in both submeasurements, our protocol correctly
predicts both outcomes.

To prove this property, notice first that if $S_1$ and $S_2$ overlap
(should they not overlap, the property is trivially true), there are two
cases: the region of overlap coincides with one of the sentences, or it
does not.  In the former case, we choose $S_2$ to be the sentence that
coincides with the region of overlap, and in the latter case, we choose
$S_1$ to be the sentence on the left and $S_2$ to be the one on the right.
With these conventions, the left boundary of the overlap region coincides
with the $Z$ that bounds the left end of $S_2$, and the right boundary of
the overlap region coincides in the former (latter) case with the $Z$ that
bounds the right end of $S_2$ ($S_1$).

To be submeasurements of the same global measurement, the two sentences
must satisfy the following basic rule: in the overlap region, sites within
a word of one sentence must be occupied in the other sentence by the same
Pauli operator or by an $I$.  Since $Z$s do not occur in words, this rule
implies that the $Z$s that bound the overlap region at either end in one
of the two sentences cannot occupy a site within a word in the other
sentence and thus must be a bounding $Z$ or a singleton $I$ in the other
sentence.  The submeasurement requirement, by itself, implies that in the
overlap region, the site of a singleton $I$ in one sentence can be
occupied by anything in the other sentence, but the available words impose
a much stronger constraint, as we now show.

Consider the left boundary of the overlap region, which is occupied by the
leftmost $Z$ in $S_2$ and by a $Z$ or a singleton $I$ in $S_1$.
Immediately to the right in both $S_1$ and $S_2$ is a word.  When one of
these words is shorter than the other, the basic rule implies that the
shorter word must be a prefix of the longer one.  A glance at the allowed
words in Eq.~(\ref{eq:primitives}) shows, however, that none is a prefix
of another. Thus $S_1$ and $S_2$ must have the same word in this first
overlap position, which is followed by a singleton~$I$ in both sentences.
Applying the same logic to this and subsequent singleton $I$s shows, as
promised, that $S_1$ and $S_2$ are identical in the overlap region, except
possibly at the boundaries.

\section{Conclusion}
\label{sec:conclusion}

Communication-assisted LHV models allow us to explore the degree of
nonlocality present in various states.  In this paper we focused on graph
states and parameterized communication-assisted LHV models by the allowed
distance of communication, where the distance between two qubits is
defined as the number of links between the corresponding nodes in its
graph. Interestingly, a simple nearest-neighbor communication protocol is
capable of yielding the global quantum-mechanical correlation for any
measurement of Pauli products on any graph state, but the submeasurements
of these global measurements are much harder to reproduce.  To replicate
the predictions of quantum mechanics for all submeasurements on any graph
state, it is necessary for the communication distance to scale as $n/6$ or
faster in the number $n$ of qubits in the graph.  Thus, using the metric
of communication distance, reproduction of all subcorrelations is a much
more difficult task than producing global correlations.

Unexpectedly, another property of interest for communication protocols
seems to be a kind of graph isomorphism symmetry, which we dubbed site
invariance.  By considering a class of two-dimensional cluster states, we
showed that, regardless of communication distance, site-invariant
communication protocols based on the local hidden variables of
Eq.~(\ref{eq:lhv}) are incapable of yielding the correct correlations for
all submeasurements on all graph states.  Nevertheless, a site-invariant
communication protocol with unlimited communication distance {\em is\/}
capable of yielding the correct correlations for all submeasurements on
all one-dimensional cluster states.  These results are notable because the
two-dimensional cluster state is a suitable resource for measurement-based
quantum computation, while the one-dimensional cluster state is not.  This
perhaps suggests a fundamental division between states such as the
two-dimensional cluster state which are sufficient for quantum computation
and states such as the GHZ and one-dimensional cluster state which are
not.

Our hope is that study of communication-assisted LHV models will lead to a
better understanding of the nature of entanglement and the apparent
nonlocality of quantum mechanics.  Already, in this paper, we have
indications that the richness of entanglement lies not in the overall
measurement result, but in measurement subcorrelations.

\acknowledgments
Research at Perimeter Institute is supported in part by the Government
of Canada through NSERC and by the Province of Ontario through MEDT.
CMC, BE, and MBE are partly supported by Army Research Office Contract
No.~W911NF-04-1-0242.  SP acknowledges support by the David and Alice
Van Buuren fellowship of the Belgian American Educational Foundation,
by the National Science Foundation under Grant No.~EIA-0086038, and by
the EU project~QIP.

\end{document}